\documentclass[10pt,letterpaper]{article}
\usepackage{opex3}
\usepackage{cite}
\usepackage{amsmath}
\usepackage{float}

\newcommand{\Nd}{$ \mathrm{Nd^{3+}} $}
\newcommand{\Er}{$ \mathrm{Er^{3+}} $}
\begin{document}

\title{Theoretical investigation of the more suitable rare earth to achieve high gain in waveguide based on silica containing silicon nanograins doped with either $ \mathbf{Nd^{3+}}$ or $\mathbf{Er^{3+}}$ ions}

\author{Alexandre Fafin,$^{*}$ Julien Cardin,$^{1}$ Christian Dufour, and Fabrice Gourbilleau}

\address{CIMAP, CNRS/CEA/ENSICAEN/UCBN \\ 6 boulevard Mar\'echal Juin, 14050 Caen cedex 4, France \\ $^{1}$julien.cardin@ensicaen.fr}

\email{$^{*}$alexandre.fafin@ensicaen.fr} 



\begin{abstract}We present a comparative study of the gain achievement in a waveguide whose active layer is constituted by a silica matrix containing silicon nanograins acting as sensitizer of either neodymium ions (\Nd) or erbium ions (\Er). By means of an auxiliary differential equation and finite difference time domain (ADE-FDTD) approach that we developed, we investigate the steady states regime of both rare earths ions and silicon nanograins levels populations as well as the electromagnetic field for different pumping powers ranging from $1$ to $10^{4}~\mathrm{mW/mm^{2}}$. Moreover, the achievable gain has been estimated in this pumping range. The \Nd doped waveguide shows a higher gross gain per unit length at 1064 nm (up to $ 30~\mathrm{dB/cm} $) than the one with \Er doped active layer at 1532 nm (up to $ 2~\mathrm{dB/cm} $). Taking into account the experimental background losses we demonstrate that a significant positive net gain can only be achieved with the \Nd doped waveguide.\end{abstract}



\section{Introduction}

The development of waveguide optical amplifiers based on rare earth (RE) doped silicon based matrix is of great interest for the semiconductor research community\cite{Daldosso2009}. Erbium ions are particularly interesting due to an optical transition at $1532~\mathrm{nm}$ which coincides with the maximum of transmission in optical glass fiber\cite{Agrawal2010}. However, in the case of erbium this transition involves the ground level, which may limit the achievable gain due to reabsorption mechanisms\cite{Lumholt1995}. To overcome this critical issue, neodymium ion has been recently proposed because its emission scheme makes it more suitable for achieving higher gain\cite{Podhorodecki2010}. In such a system, the amplification is commonly achieved through RE level population inversion by an appropriate optical pumping. One main drawback of RE ions is their low absorption cross section. However this can be overcome by the use of sensitizers that are characterized by a larger absorption cross section. Those sensitizers have shown an efficient transfer of energy to RE ions in their vicinity. Several sensitizers of RE have been proposed in literature. Polman \textit{et al} \cite{Polman2004} shows the sensitization of \Er ions by different kinds of sensitizers such as ytterbium ions, metal ions and silicon nanograins (Si-ng). The work of MacDonald \textit{et al} \cite{MacDonald2006} presents the sensitization of \Nd ions by Si-ng.

In  this paper we present a comparative study of waveguides with an active layer containing Si-ng and doped either with erbium or neodymium ions. The typical composition and structure of such waveguides is presented in section \ref{description waveguide}. The electromagnetic field and level populations of Si-ng and RE ions have been computed using an algorithm published in a previous paper \cite{Fafin2013} and briefly detailed in section \ref{calculation method}.

In section \ref{rare earth}, we describe levels populations equations associated with Si-ng, erbium ions and neodymium ions. For the two RE ions, due to their different transition time properties, two particular ways of calculation will be detailed. In section \ref{results}, we present for both RE population inversions, population map and optical gain as a function of optical pump power. We conclude by the comparison of the optical gain of two waveguides doped either with erbium or neodymium ions as a function of the pump power.

\subsection{General description of the waveguide}
\label{description waveguide}
\begin{figure}[h!]
	\centering
	\includegraphics[width=0.7\textwidth]{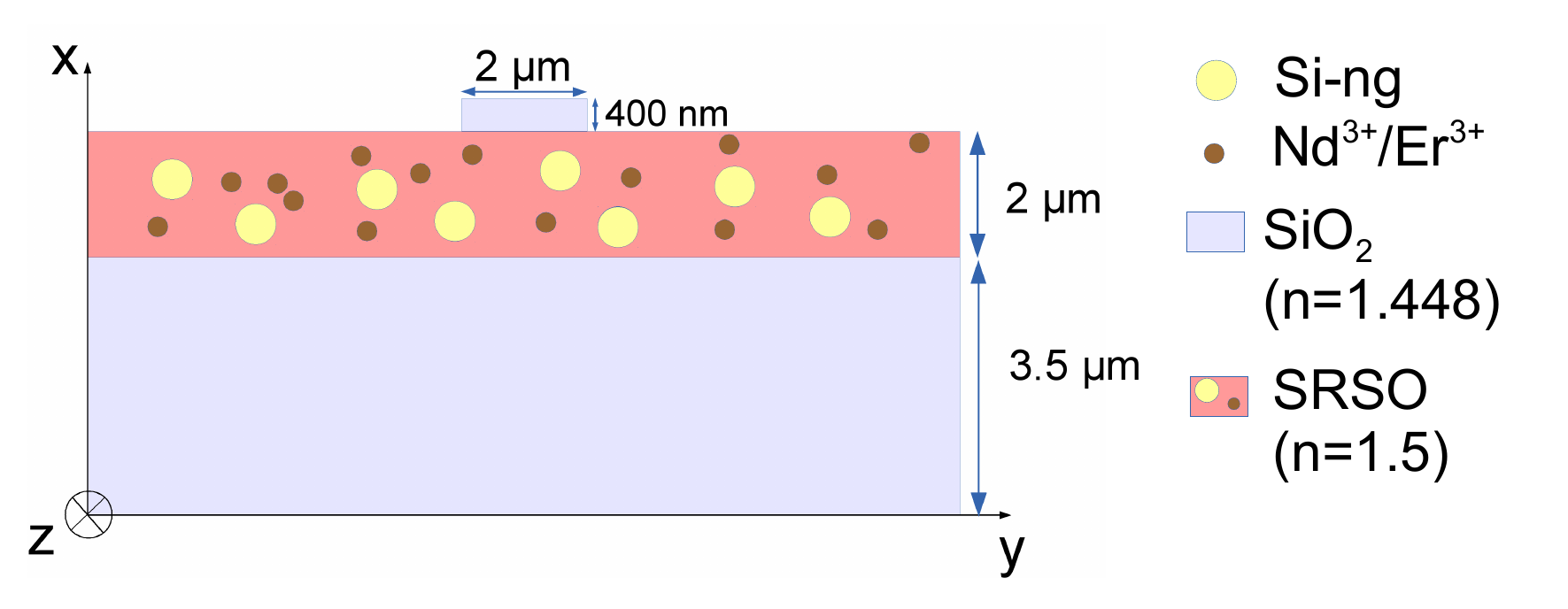}
	\caption{Transverse section view of the waveguide constituted by bottom and strip cladding layers of silica surrouding the active layer constituted by silicon rich silicon oxide (SRSO) matrix doped with silicon nanograins (Si-ng) and \Nd or \Er ions.}
	\label{waveguide}
\end{figure}

The waveguide is composed of three layers (Fig. \ref{waveguide}). The bottom cladding layer is composed of pure silica. In order to ensure optical confinement of modes, this layer is about 5 to 8 $\mathrm{\mu}$m thick in a typical experimental waveguide doped either with \Nd ions \cite{Pirasteh2012} or \Er ions \cite{Daldosso2006}. In this modeling method the thickness of bottom cladding layer was taken equal to $3.5~\mathrm{\mu m}$ in order to limit the use of memory. The $2~\mathrm{\mu m}$ active layer constituted of silicon rich silicon oxide (SRSO) contains Si-ng and RE ions. A pure silica strip layer is stacked on the top of the SRSO layer. The static refractive index (\textit{i.e.} refractive index which remains constant with wavelength) of the active layer ($1.5$) has been chosen greater than the one of the strip and bottom cladding layers ($1.448$) to ensure the guiding conditions. 

In order to investigate which is the most suitable RE between \Nd and \Er for achieving high gain, both waveguides are pumped continuously (CW) by the propagation in the active layer of a pump mode at 488 nm. A signal mode is co-propagated in the active layer in order to investigate the achievability of amplification by stimulated emission. This signal corresponds to a transition occurring between electronics levels of RE, either at 1532 nm for erbium ions or at 1064 nm for neodymium ions\cite{Barnes1991,Serqueira2006}. The waveguide dimensions are identical for erbium or neodymium ions and according to the experimental conditions we propagate the fundamental transverse electric mode ($\mathrm{TE_{00}}$) for the pump and signal along z direction. The calculation and the injection of mode profiles in ADE-FDTD method for all wavelengths considered here are described in our previous paper\cite{Fafin2013}.

\subsection{Calculation method}
\label{calculation method}
The electromagnetic fields ($ \mathbf{E},\mathbf{H}$) and Poynting vector ($\mathbf{R}= \mathbf{E}\times\mathbf{H}$) as well as RE and Si-ng populations in steady state are computed by the algorithm described by Fafin et al\cite{Fafin2013}. This algorithm is based on finite-difference time-domain method (FDTD) and auxiliary differential equations (ADE). The FDTD method consists in the discretization in time and space of the electromagnetic fields\cite{Taflove1995} according to the K. Yee algorithm\cite{Yee1966}. The ADE makes the link between electromagnetic fields and absorption and emission processes by the use of a set of polarisation densities $ \mathbf{P}_{ij}$ following a Lorentz electronic oscillator model\cite{Taflove1995,Nagra1998}. A typical polarisation equation (Eq. (\ref{polarisation})) between two level populations $i$ and $j$ is described below:
\begin{equation}
\label{polarisation}
	\frac{\partial^2 \mathbf{P}_{ij}}{\partial t^2} + \Delta \omega_{ij} \frac{\partial \mathbf{P}_{ij}}{\partial t} + \omega_{ij}^2 \mathbf{P}_{ij} = \kappa_{ij} (N_i-N_j) \mathbf{E}
\end{equation}
where $ \Delta \omega_{ij} $ is the transition linewidth including radiative, non-radiative and de-phasing processes\cite{Dufour2011}, and $\omega_{ij}$ is the resonance frequency of this transition. $\kappa_{ij}$ used in \cite{Fafin2013} depends on the transition lifetime $\tau_{ij}$ and on the optical index n.

The time evolution of levels populations for each emitter (RE, Si-ng) is described by a rate equation which depends on polarisation densities of considered transitions, lifetimes, transfer coefficient and levels populations. Since in visible wavelength range, the electromagnetic field has a characteristic time of the order of $10^{-15}$ s and the levels populations of emitters have characteristic lifetimes as long as a few milliseconds \cite{Kenyon2002,Ainslie1989}, a classical ADE-FDTD calculation is impossible in a reasonable time\cite{Fafin2013}. Indeed with the classical ADE-FDTD method the equations of populations are calculated simultaneously with the electromagnetic field leading to about $10^{15}$ iterations. We have recently overcome this multiscale issue \cite{Fafin2013} after splitting the classical ADE-FDTD single loop into two interconnected loops separating short time and long time processes. The electromagnetic fields and polarisation densities are calculated in the short time loop and the rate equations in the long time one. This method allows us to reduce drastically the number of iterations from $10^{15}$ to $10^5$ and consequently reduces the calculation time to more reasonable duration (7 days for the present waveguide at bi-processors quad-core Intel Nehalem EP @ 2.8 GHz). Moreover, in order to minimize phase velocity error and velocity anisotropy errors inherent to the FDTD method, the algorithm was set up with the time and space steps introduced in our previous paper \cite{Fafin2013}. 

\section{Description of populations}
\label{rare earth}
In this section we describe the excitation mechanism of the erbium and neodymium ions and the numerical calculation of the levels populations at the steady state.

\subsection{Silicon nanograins}

\begin{figure}[h!]
	\includegraphics[width=0.49\textwidth]{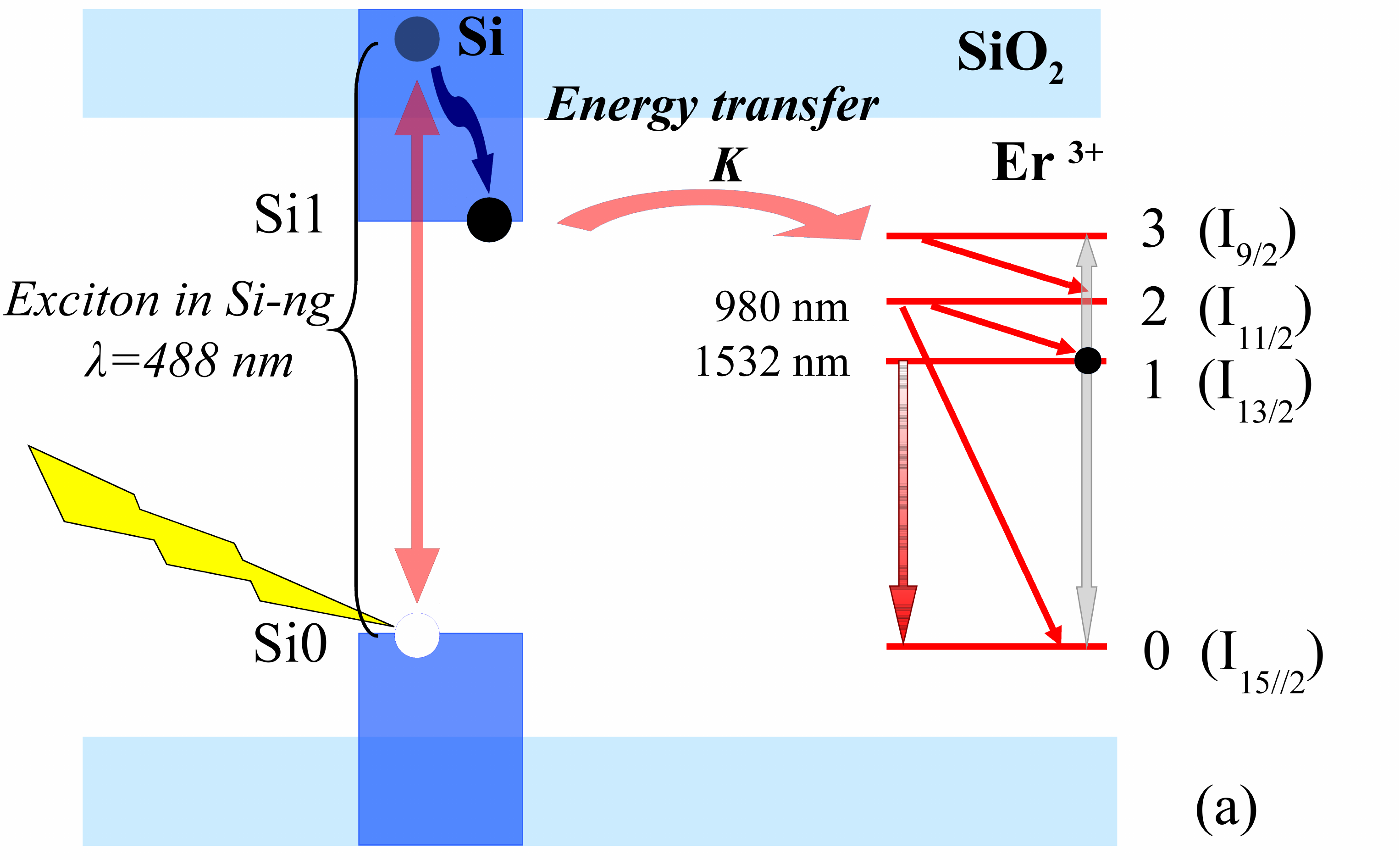}
	\includegraphics[width=0.49\textwidth]{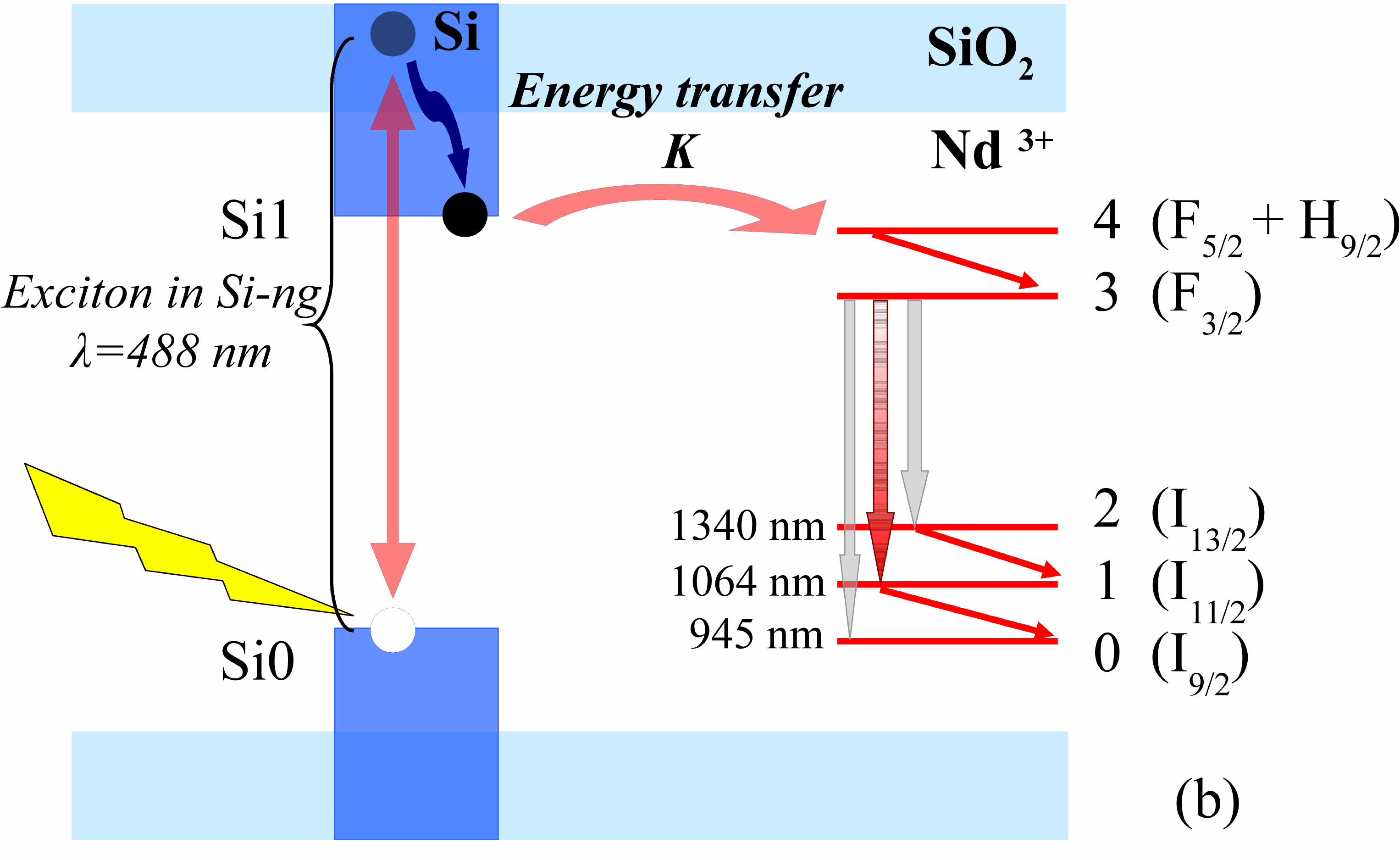}
	\caption{Excitation mechanism of (a) erbium ions and (b) neodymium ions}
	\label{schema excitation}
\end{figure}
  
We model silicon nanograins (Si-ng) as a two levels system (Fig. \ref{schema excitation}) where the ground and excited levels populations (respectively $\mathrm{N_{Si0}}$ and $\mathrm{N_{Si1}}$) are given by the rate equations Eq. (\ref{eq_si_0}) and Eq. (\ref{eq_si_1}). Due to a low probability of multi-exciton generation in a single Si-ng \cite{Govoni2012}, we assume the excitation of one single exciton by Si-ng, therefore the Si-ng population will correspond to the exciton population. After non-radiative transitions in the conduction band, the exciton may either radiatively recombine or excite an emitter in its vicinity. This energy transfer may occur if the energy gap between conduction and valence band of Si-ng matches the energy gap between the RE fundamental level and an upper level leading to a possible emission. According to literature \cite{Priolo2001} the lifetime of excited level $\mathrm{N_{Si1}}$ is chosen at $ \tau_{Si_{10}}\vert^r_{nr}=50~\mathrm{\mu s} $. Since few papers studied the energy transfer coefficient in Si-ng and erbium ions \cite{Pacifici2003,Toccafondo2008} and to our knowledge no study was made on this coefficient in Si-ng and neodymium, this energy transfer coefficient between Si-ng and both RE is assumed identical and taken equal to $ K=10^{-14}~\mathrm{cm^3/s} $. We took the same concentration of Si-ng equal to $ 10^{19}~\mathrm{cm^{-3}}$ for both active layers doped either with \Er or \Nd ions. In order to simulate a realistic absorption and emission cross section equal to $ 10^{-16}~\mathrm{cm^2} $ the linewidth $ \Delta \omega_{ij} $ and the number of polarizations $ \mathrm{N_p}$ are respectively fixed to $ 10^{14}~\mathrm{rad/s} $ and 2756 according to the method explained in \cite{Fafin2013}. These parameters are reported in Table \ref{silicon parameters}.  

\begin{equation}
\label{eq_si_0}
\dfrac{dN_{Si_1}\left(t\right)}{dt}=+\frac{1}{\hbar\omega_{Si_{10}}}\textbf{E}\left(t\right)\frac{d\textbf{P}_{Si_{10}} \left(t\right)}{dt}-\dfrac{N_{Si_1}\left(t\right)}{\tau_{Si_{10}}\vert _{nr}^{r}}-KN_{Si_1}\left(t\right)N_{0}\left(t\right)
\end{equation}

\begin{equation}
\label{eq_si_1}
\dfrac{dN_{Si_0}\left(t\right)}{dt}=-\frac{1}{\hbar\omega_{Si_{10}}} \textbf{E}\left(t\right)\frac{d\textbf{P}_{Si_{10}}\left(t\right)}{dt}+\dfrac{N_{Si_1}\left(t\right)}{\tau_{Si_{10}}\vert _{nr}^{r}}+KN_{Si_1}\left(t\right)N_{0}\left(t\right)
\end{equation}

\renewcommand{\arraystretch}{1.3}
\begin{table}
	\footnotesize
	\caption{Parameters levels of silicon nanograins}
	\centering
	\begin{tabular}{ccccc}
		\hline
		$j \rightarrow i$ & Lifetime (s) & $ \omega_{ij} (10^{15}~\mathrm{rad/s}) $ & $ \Delta \omega_{ij} (10^{14}~\mathrm{rad/s}) $ & $ N_p $\\
		\hline
		$ 1\rightarrow0 $ & $ 50.10^{-6} $ & 3.682 & 1 & 2756 \\
	\end{tabular}
	\label{silicon parameters}
\end{table}

\subsection{Erbium ions}
Erbium ions are modelled by four levels: 0 ($^4I_{15/2}$), 1 ($^4I_{13/2}$), 2 ($^4I_{11/2}$) and 3 ($^4I_{9/2}$). We consider two non-radiative transitions 3$\rightarrow$2, 2$\rightarrow$1 and one radiative transition from level 1 to level 0 at 1532 nm. The emission cross section of this transition is equal to $ 6\times10^{-21}\mathrm{cm^{-2}} $\cite{Pacifici2003,Toccafondo2008} and corresponds to a linewidth $ \Delta \omega_{10} $ equal to $ 0.15\times10^{15}~\mathrm{rad/s} $ with one polarisation ($N_p=1$)\cite{Fafin2013}. Moreover there is an up-conversion process from level 1 to level 0 and 3 which can be modelled by a coefficient $ C_{up}=5\times10^{-17} ~\mathrm{cm^3/s}$. The concentration of erbium ions is equal to $ 10^{20}~\mathrm{cm^{-3}} $. The time evolution of \Er levels populations is described by the rate equations Eq. (\ref{eq_er_3}) to Eq. (\ref{eq_er_0}). All parameters of erbium ions transitions are taken from \cite{Pacifici2003,Toccafondo2008} and reported in Table \ref{erbium parameters}. 

\begin{align}
\label{eq_er_3}
\dfrac{dN_{3}\left(t\right)}{dt}=&-\dfrac{N_{3}\left(t\right)}{\tau_{32}\vert_{nr}}+KN_{Si_1}\left(t\right)N_{0}\left(t\right)+C_{up} N_1^2 \\
\label{eq_er_2}
\dfrac{dN_{2}\left(t\right)}{dt}=&+\dfrac{N_{3}\left(t\right)}{\tau_{32}\vert_{nr}}-\dfrac{N_{2}\left(t\right)}{\tau_{21}\vert_{nr}}-\dfrac{N_{2}\left(t\right)}{\tau_{20}\vert_{nr}} \\
\label{eq_er_1}
\dfrac{dN_{1}\left(t\right)}{dt}=&+\frac{1}{\hbar\omega_{10}}\textbf{E}\left(t\right)\frac{d\textbf{P}_{10}\left(t\right)}{dt}+\dfrac{N_{2}\left(t\right)}{\tau_{21}\vert_{nr}}-\dfrac{N_{1}\left(t\right)}{\tau_{10}\vert_{nr}^r}-2 C_{up} N_1^2 \\
\label{eq_er_0}
\dfrac{dN_{0}\left(t\right)}{dt}=& -\frac{1}{\hbar\omega_{10}}\textbf{E}\left(t\right)\frac{d\textbf{P}_{10}\left(t\right)}{dt} \\ \nonumber &+\dfrac{N_{2}\left(t\right)}{\tau_{20}\vert_{nr}}+\dfrac{N_{1}\left(t\right)}{\tau_{10}\vert_{nr}^r}-KN_{Si_1}\left(t\right)N_{0}\left(t\right)+C_{up} N_1^2
\end{align}

\begin{table}[h!]
	\footnotesize
	\caption{Parameters levels of erbium ions}
	\centering
	\begin{tabular}{ccccc}
		\hline
		$j \rightarrow i$  & 3$\rightarrow$2   & 2$\rightarrow$1     & 2$\rightarrow$0     & 1$\rightarrow$0 \\
		\hline
		Lifetime (s)& $0.1\times10^{-6}$   & $2.4\times10^{-6}$& $ 710\times10^{-6} $& $ 8.5\times10^{-3} $\\
		$\omega_{ij}$ ($10^{15}~\mathrm{rad/s}$) &             &   1.23              &                     &   \\
		$\Delta \omega_{ij}$ ($10^{15}~\mathrm{rad/s}$)&       &   0.15              &                     &    \\
		$ N_p $	&       &   1              &                     &  
	\end{tabular}
	\label{erbium parameters}
\end{table}

\subsection{Neodymium ions}
Neodymium ions are modelled by five levels: 0 ($^4I_{9/2}$), 1 ($^4I_{11/2}$), 2 ($^4I_{13/2}$), 3 ($^4F_{3/2}$) and 4 ($^4F_{5/2}+^2H_{9/2}$). We consider three non-radiative transitions 4$\rightarrow$3, 2$\rightarrow$1, 1$\rightarrow$0 and three radiative transitions 3$\rightarrow$2 at 1340 nm, 3$\rightarrow$1 at 1064 nm and 3$\rightarrow$0 at 945 nm.  The emission cross section of these transitions is equal to $ 10^{-19}\mathrm{cm^{-2}} $\cite{Siegman1986} and correspond to linewidths $ \Delta \omega_{ij} $ reported in Table \ref{neodymium parameters}. The up-conversion coefficient $C_{up}$ ranging from $ 1\times10^{-17}$ to $ 5\times10^{-17}~\mathrm{cm^3/s} $ was found in \cite{Oliveira2008,Dalfsen2010}. This value leads to an equivalent lifetime $\tau_{up}=\frac{1}{C_{up}N_1}$ that remains 10 times larger than the longest level lifetime of \Nd ions which allows neglecting the up-conversion process\cite{Fafin2013}. The concentration of neodymium ions is equal to $ 10^{20}~\mathrm{cm^{-3}} $. The time evolution of \Nd levels populations is described by the rate equations Eq. (\ref{eq_nd_4}) to Eq. (\ref{eq_nd_0}). Parameters of neodymium ions transitions are taken from \cite{Serqueira2006,Siegman1986} and reported in Table \ref{neodymium parameters}. 

\begin{align}
\label{eq_nd_4}
\frac{dN_{4}\left(t\right)}{dt}=&-\frac{N_{4}\left(t\right)}{\tau_{43}\vert_{nr}}+KN_{Si_1}\left(t\right)N_{0}\left(t\right) \\
\label{eq_nd_3}
\frac{dN_{3}\left(t\right)}{dt}=&+\frac{1}{\hbar\omega_{30}}\textbf{E}\left(t\right)\frac{d\textbf{P}_{30}\left(t\right)}{dt}+\frac{1}{\hbar\omega_{31}}\textbf{E}\left(t\right)\frac{d\textbf{P}_{31}\left(t\right)}{dt}+\frac{1}{\hbar\omega_{32}}\textbf{E}\left(t\right)\frac{d\textbf{P}_{32}\left(t\right)}{dt} \\ \nonumber &+\frac{N_{4}\left(t\right)}{\tau_{43}\vert_{nr}}-\frac{N_{3}\left(t\right)}{\tau_{30}\vert_{nr}^r}-\frac{N_{3}\left(t\right)}{\tau_{31}\vert_{nr}^r}-\frac{N_{3}\left(t\right)}{\tau_{32}\vert_{nr}^r} \\
\label{eq_nd_2}
\frac{dN_{2}\left(t\right)}{dt}=&-\frac{1}{\hbar\omega_{32}}\textbf{E}\left(t\right)\frac{d\textbf{P}_{32}\left(t\right)}{dt}+\frac{N_{3}\left(t\right)}{\tau_{32}\vert_{nr}^r}-\frac{N_{2}\left(t\right)}{\tau_{21}\vert_{nr}} \\
\label{eq_nd_1}
\frac{dN_{1}\left(t\right)}{dt}=&-\frac{1}{\hbar\omega_{31}}\textbf{E}\left(t\right)\frac{d\textbf{P}_{31}\left(t\right)}{dt}+\frac{N_{3}\left(t\right)}{\tau_{31}\vert_{nr}^r}-\frac{N_{1}\left(t\right)}{\tau_{10}\vert_{nr}}+\frac{N_{2}\left(t\right)}{\tau_{21}\vert_{nr}} \\
\label{eq_nd_0}
\frac{dN_{0}\left(t\right)}{dt}=&-\frac{1}{\hbar\omega_{30}}\textbf{E}\left(t\right)\frac{d\textbf{P}_{30}\left(t\right)}{dt}+\frac{N_{3}\left(t\right)}{\tau_{30}\vert_{nr}^r}+\frac{N_{1}\left(t\right)}{\tau_{10}\vert_{nr}}-KN_{Si_1}\left(t\right)N_{0}\left(t\right)
\end{align}

\begin{table}[h!]
	\footnotesize
	\caption{Parameters levels of neodymium ions}
	\centering
	\begin{tabular}{ccccccc}
		\hline
		$ j \rightarrow i$  & 4$\rightarrow$3 & 3$\rightarrow$2 & 3$\rightarrow$1 & 3$\rightarrow$0 & 2$\rightarrow$1 & 1$\rightarrow$0 \\
		\hline
		Lifetime (s)& $230\times10^{-12}$  & $ 1000\times10^{-6} $   & $200\times10^{-6}$   & $250\times10^{-6}$ & $970\times10^{-12}$ & $510\times10^{-12}$  \\
		$\omega_{ij}$ ($10^{15}~\mathrm{rad/s}$) &   &   1.34   & 1.77 & 1.99 &  &  \\
		$\Delta \omega_{ij}$ ($10^{15}~\mathrm{rad/s}$)&  & 0.67& 0.18 & 0.11&  &  \\
		$ N_p $ &  & 1& 1 & 1&  &
	\end{tabular}	
	\label{neodymium parameters}
\end{table}

\subsection{Difference in calculation method of populations}
We aim at calculating levels populations in steady states in a fast and convenient manner. Consequently, the preferred method would be an analytical calculation of populations by setting $ dN_i/dt=0$. However, this is only applicable in case of neodymium levels equations. Moreover, the up-conversion term in erbium rate equations leads to equations that are hardly analytically solvable. In that case,  the steady states of population levels are reached by a finite difference method. This calculation was possible using a reasonable time step of $0.01~\mathrm{\mu s}$ ten times lower than the shortest lifetime ($0.1~\mathrm{\mu s}$) considered in this model. The calculation time is then no longer negligible but does not rise up significantly the global calculation time of our ADE-FDTD method.

\section{Results and discussion}
\label{results}
The solution of Eq. (\ref{eq_si_0}) to Eq. (\ref{eq_nd_0}) gives the levels populations in their steady states. We define for $i \rightarrow j$ transition the population inversion (section \ref{populations inversion}) as the ratio ($N_i-N_j$) over the total population number ($N_{tot}=10^{20}~\mathrm{at/cm^{3}}$ for RE or $ N_{Si_{tot}}=10^{19}~\mathrm{at/cm^{3}} $ for Si-ng). We deduced also the optical gain (section \ref{optical gain}) at $1064~\mathrm{nm}$ for \Nd and $ 1532~\mathrm{nm} $ for \Er. These values are computed for a pump power ranging from $1$ to $10^{4}~\mathrm{mW/mm^{2}}$.

\subsection{Populations inversion}
\label{populations inversion}
\begin{figure}[h!]
	\centering
	\includegraphics[width=0.49\textwidth]{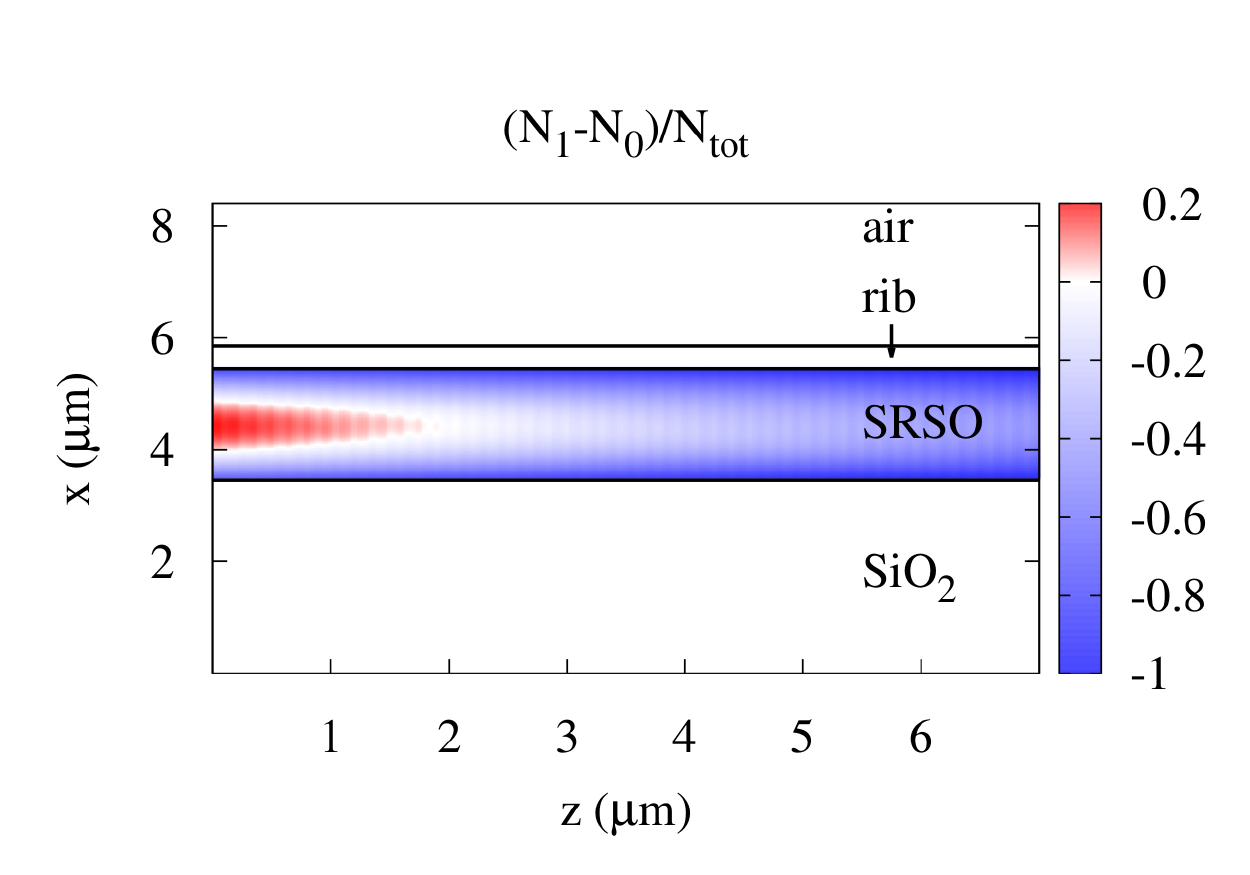}
	\includegraphics[width=0.49\textwidth]{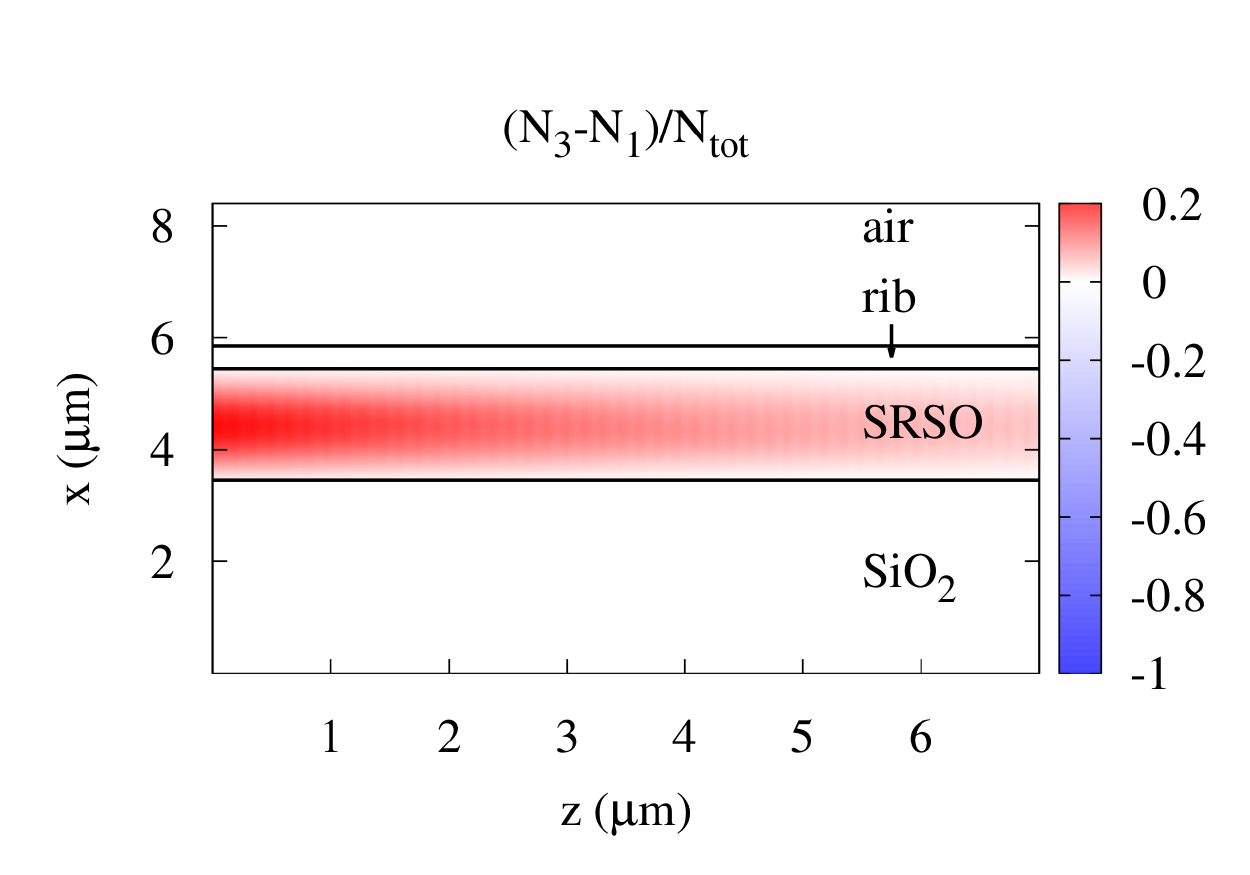}
	\caption{Population inversion along the direction of propagation for the erbium ions (on the left) and the neodymium ions (on the right) for a pump power equal to $1000~\mathrm{mW/mm^{2}} $}
	\label{populations RE}
\end{figure}

We present in this section the spatial distribution of population inversion in waveguides doped either with erbium or neodymium ions for a pump power of $ 1000~\mathrm{mW/mm^{2}} $ in a longitudinal section view along the propagation axis (Fig. \ref{populations RE}). For both waveguides the plots show a decrease of population inversion with direction of propagation which can be attributed to the coupling between rare earth ions and silicon nanograins.

For erbium ions, the population inversion remains positive over a length of $1.5~\mathrm{\mu m}$. Beyond this length, the population inversion becomes negative witnessing the threshold effect occurring with three-level system. 

For neodymium ions the population inversion remains positive along the whole structure. Indeed, in this four-level system, the level 1 is depopulated quickly to the ground level leading to $N_3 >> N_1$.

\begin{figure}[h!]
	\centering
	\includegraphics[width=0.49\textwidth]{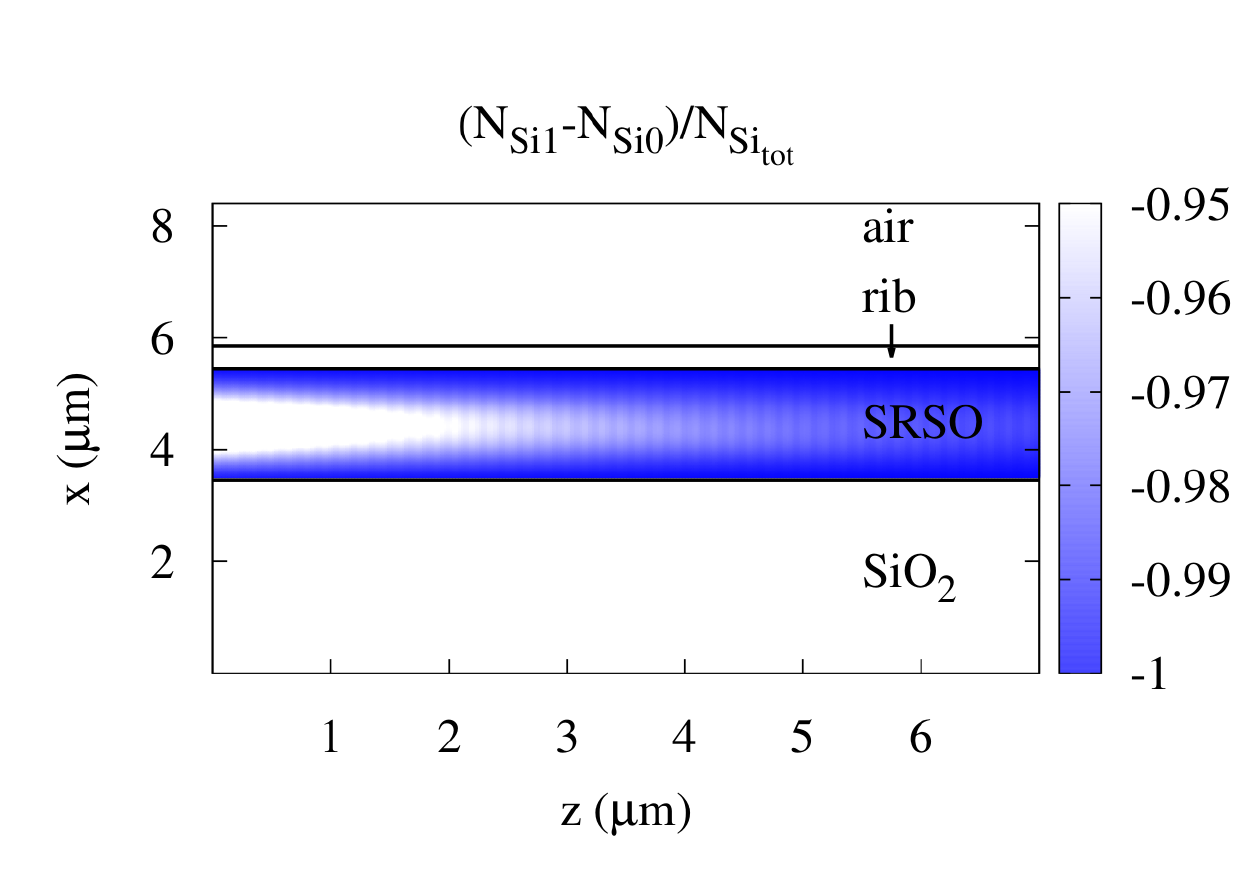}
	\includegraphics[width=0.49\textwidth]{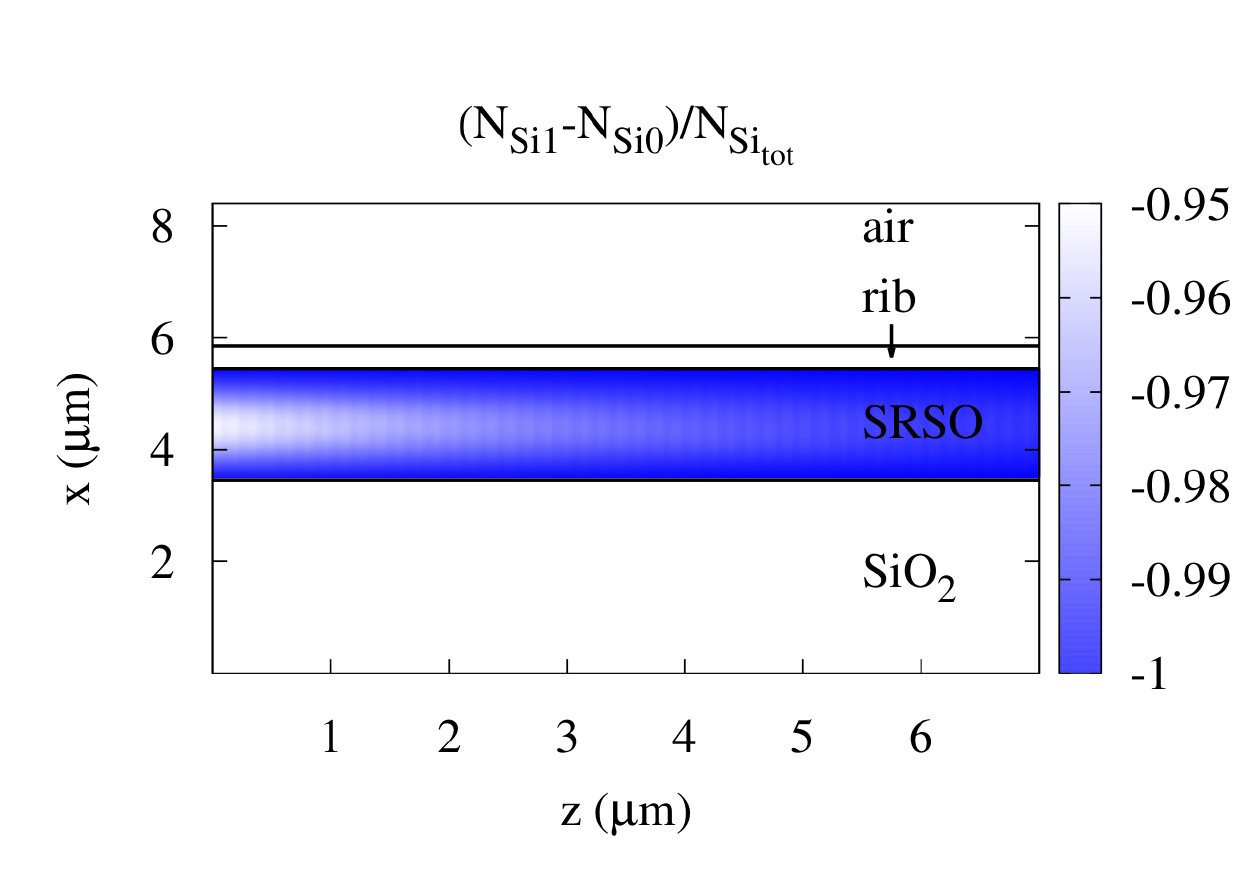}
	\caption{Population inversion of Si-ng along the direction of propagation in the case of erbium ions (on the left) and neodymium ions (on the right) for a pump power equal to $1000~\mathrm{mW/mm^{2}} $}
	\label{populations Si}
\end{figure}

We now consider the population inversion distribution of Si-ng presented in longitudinal section view in Fig. \ref{populations Si} for both RE. This decrease of population inversion of Si-ng shows an identical behaviour with direction of propagation as the one observed with RE. This is characteristic of the pump strong absorption due to the presence of the nanograins as shown in our previous paper \cite{Fafin2013}. 

\begin{figure}[h!]
	\centering
	\includegraphics[width=0.8\textwidth]{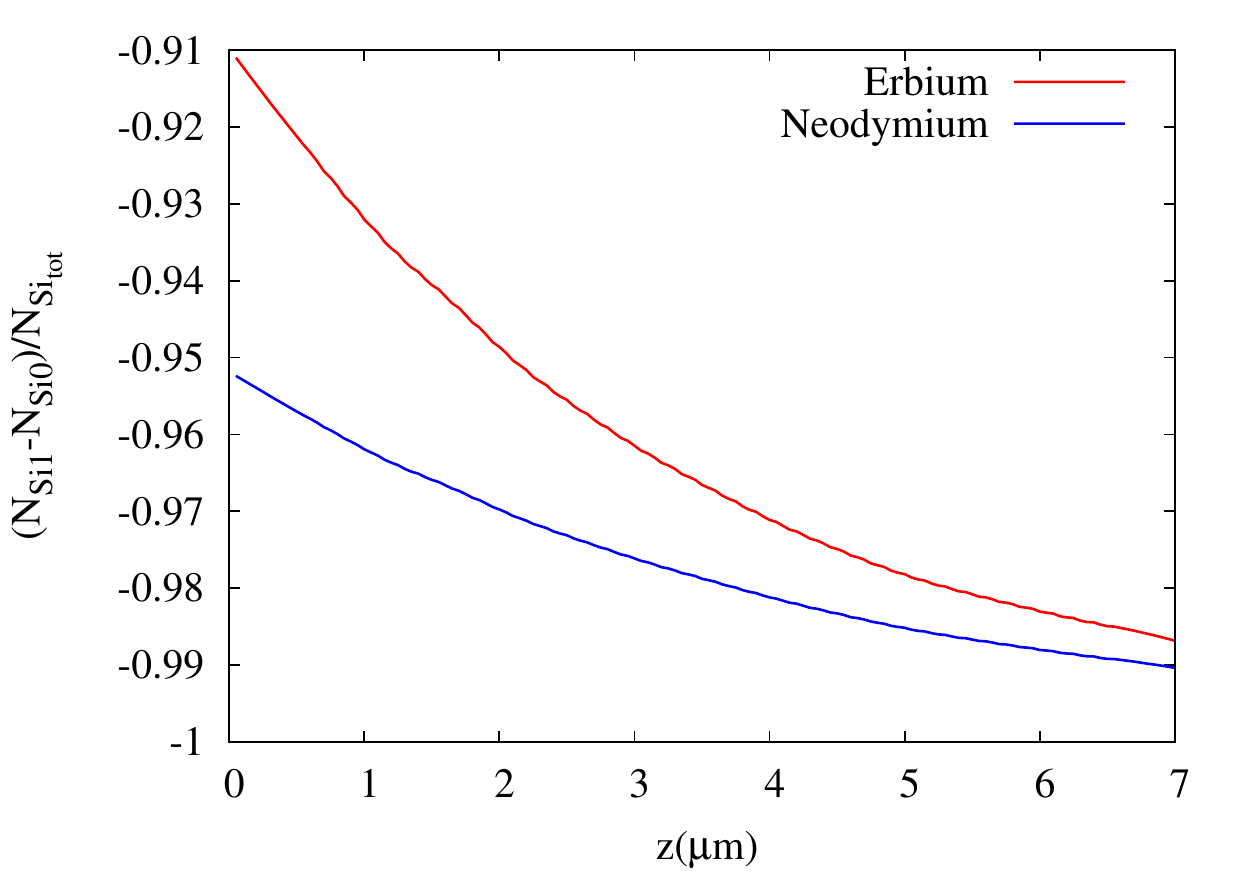}
	\caption{Population inversion of Si-ng along the direction of propagation in the case of erbium and neodymium recorded at $x=4.5~\mathrm{\mu m}$ and $y= 8.55~\mathrm{\mu m}$ (center of the XY section of the active layer) for a pump power equal to $1000~\mathrm{mW/mm^{2}} $}
	\label{populations Si z}
\end{figure}

From Fig. \ref{populations Si} we extract a particular set of values of the population inversion of Si-ng along the direction of propagation recorded at $x=4.5~\mathrm{\mu m}$ and $y= 8.55~\mathrm{\mu m}$ (center of the XY section of the active layer) (Fig. \ref{populations Si z}). On this figure, we observe along the whole length of the waveguide a larger Si-ng population inversion with erbium than with neodymium. Since the Si-ng modeling is the same with both RE, the  population inversion difference is due to each specific Si-ng/RE interaction. This interaction is governed by the transfer coefficient K, which is the same for both RE, but also by specific transitions lifetimes of each RE. The larger population inversion observed for erbium than with neodymium is consequently due to the difference of lifetimes between these RE (Table \ref{erbium parameters} and \ref{neodymium parameters}). This observation leads us to the conclusion that RE excitation occurring from Si-ng is more efficient in the case of neodymium than in the case of erbium ions due to the specific time dynamics of transitions in different RE.

\begin{figure}[h!]
	\centering
	\includegraphics[width=0.8\textwidth]{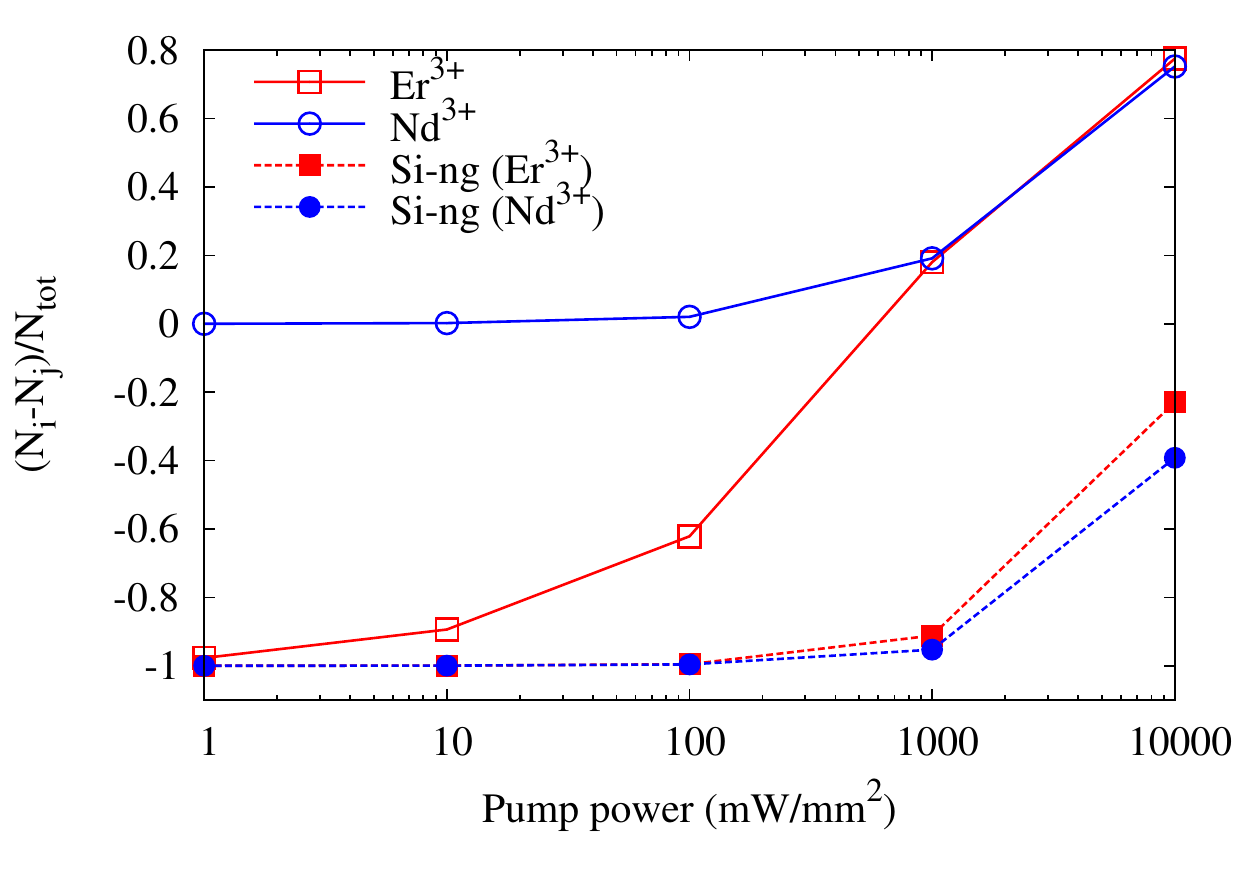}
	\caption{Population inversion for neodymium, erbium ions and silicon nanograins divided by the rare earth ions concentration ($10^{20}~\mathrm{at/cm^{3}}$) recorded at $x=4.5~\mathrm{\mu m}$ and $y= 8.55~\mathrm{\mu m}$ (center of the XY section of the active layer) and z=0 (beginning of the waveguide)}
	\label{population_inversion}
\end{figure}

Figure \ref{population_inversion} shows the influence of pump power on population inversion for the erbium and the neodymium ions as well as for the Si-ng. We observe that the population inversion occurs above a threshold pump power ($600~\mathrm{mW/mm^{2}}$) for erbium. This behaviour is typical of a three-level system. For neodymium ions, we observe a positive population inversion for the whole pump power range which is characteristic of a four-level system. For high pump power ($1000~\mathrm{mW/mm^{2}}$), both populations inversions reach a comparable value witnessing the saturation of the excitation mechanism.

Finally, we plot (dotted lines) the population inversion of Si-ng as a function of pump power. Whatever the pump power, this inversion is higher for erbium than for neodymium ions. This evidences a more efficient transfer from Si-ng to neodymium ions than to erbium ions.

\subsection{Optical gain}
\label{optical gain}
\begin{figure}[h!]
	\centering
	\includegraphics[width=0.8\textwidth]{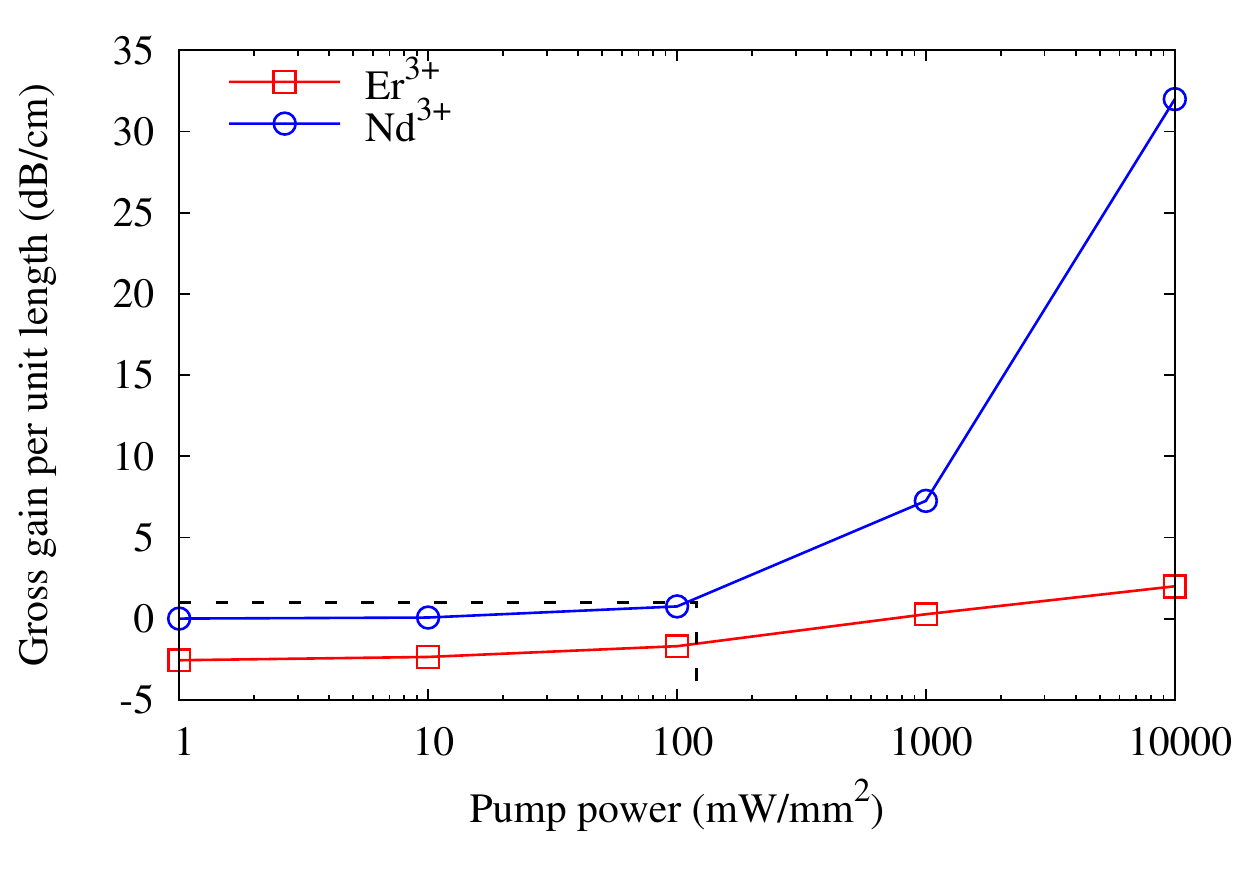}
	\caption{Local gross gain per unit length at the center of the active layer and in the beginning of the waveguide as a function of the pumping power for a waveguide doped with \Nd (open circle) and a waveguide doped with \Er (open square) recorded at $x=4.5~\mathrm{\mu m}$ and $y= 8.55~\mathrm{\mu m}$ (center of the XY section of the active layer) and $ z=0 $ (beginning of the waveguide). Losses found by Pirasteh \textit{et al} \cite{Pirasteh2012} are represented (dashed line).}
	\label{figure gain}
\end{figure}

From population levels $N_i(x,y,z)$ we deduced the local gross gain per unit length $g_{dB/cm}$ (Eq. (\ref{gain})) at the signal wavelength:
\begin{equation}
g_{dB/cm}(x,y,z)=\frac{10}{ln10}\left( \sigma_{em}N_{high}(x,y,z) - \sigma_{abs}N_{low}(x,y,z)\right)
\label{gain}
\end{equation}
where $ N_{high} $ and $ N_{low} $ are respectively the higher and lower levels of the considered transition and $ \sigma_{abs} $ and $ \sigma_{em}$ are absorption and emission cross sections. For erbium ions, we make the link $N_{high} = N_{1}$ and $N_{low} = N_{0}$ and for neodymium ions $N_{high} = N_{3}$ and $N_{low} = N_{1}$ and we assume equal emission and absorption cross sections for one RE. The local gross gain per unit length recorded at $x=4.5~\mathrm{\mu m}$ and $y= 8.55~\mathrm{\mu m}$ (center of the XY section of the active layer) and $ z=0 $ (beginning of the waveguide) is plotted in Fig. \ref{figure gain}.

For \Er doped waveguide, we find that, above a threshold pump power of $ 1550~\mathrm{mW/mm^{2}} $, a positive gross gain is reached which increases up to 
$ 2~\mathrm{dB/cm} $ for the highest pump power simulated in our study. In order to estimate the net gain, we must account for the background losses such as those found by Navarro-Urrios \textit{et al} \cite{Navarro-Urrios2011} on comparable samples ($ 3.0~\mathrm{dB/cm} $ at $1532~\mathrm{nm}$). We can conclude that it is not possible to reach a positive net gain in this range of pump power. However Navarro-Urrios \textit{et al} found a positive net gain equal to $ 0.3~\mathrm{dB/cm} $, this small difference with our modelling result may be explained by higher pump power that was used in the experiment ($4.10^5~\mathrm{mW/mm^{2}}$ to $ 6.10^6~\mathrm{mW/mm^{2}}$). 

For \Nd doped waveguide, we find that the optical gain remains positive over the whole power range. It increases up to $ 30~\mathrm{dB/cm} $ for the highest pump power of $ 10^{4}~\mathrm{mW/mm^{2}} $. We can also estimate a net gain taking into account background losses of $0.8~\mathrm{dB/cm}$ found by Pirasteh \textit{et al} in a similar system\cite{Pirasteh2012}. Figure \ref{figure gain} shows  that for a pump power above $ 130~\mathrm{mW/mm^{2}} $ the losses (dashed line) can be compensated leading to a net gain. 

The gross gain per unit length obtained for \Nd remains higher that the one obtained for \Er whatever the pump power range. The gross gain per unit length difference (between \Nd and \Er doped waveguide) is all the higher as the pump power increases. This feature may be due to the difference in Si-ng/RE transfer efficiency (section \ref{populations inversion}) linked to the levels dynamics as well as to the difference in absorption/emission cross section ($\sigma_{Er}=6\times10^{-21}~\mathrm{cm^2}$ against $\sigma_{Nd}=1\times10^{-19}~\mathrm{cm^2}$).
The three categories of commercially available optical amplifier on C-band (EDFA, EDWA and SOA) present a gain level about 20 to 25 dB with a working length ranging from cm to few meters for EDFA and for a power consumption mainly dedicated to optical or electrical pumping of about few watts \cite{Zimmerman2004}. Since we found a low gain and a short length of positive population inversion by modeling of \Er doped based waveguide on the broad range of pump power. We conclude to the impossibility of achievement of an optical amplifier with this configuration of co-propagating pump and signal which would compete with commercially available systems. The gross gain per unit length reachable with \Nd, about one order of magnitude larger than the one obtained with \Er, could lead a significant amplification. To our knowledge, there is no commercially available comparable optical amplifier based on Nd3+ emission bands. However, \Nd doped Aluminum oxide channel waveguide amplifiers developed by Yang \textit{et al} show a maximal internal gain of 6 dB/cm for a pump power of 45 mW\cite{Dalfsen2010}.

\section{Conclusion}

Our algorithm based on ADE-FDTD method previously used for modeling steady states of fields, population levels and gross gain per unit length for \Nd doped waveguide has been extended with success to the case of erbium ions doped waveguide. 

We have demonstrated that the neodymium ions are more suitable than the erbium ions to obtain a net positive gain per unit length in silica based waveguide containing silicon nanograins. The theoretical maximum gross gain per unit length of $ 2~\mathrm{dB/cm} $ at 1532 nm ($ 10^{4}~\mathrm{mW/mm^{2}} $) does not compensate background losses experimentally estimated to $ 3~\mathrm{dB/cm} $. On the contrary, the use of neodymium ions leads to a gross gain per unit length of $ 30~\mathrm{dB/cm} $ at 1064 nm ($ 10^{4}~\mathrm{mW/mm^{2}} $). Moreover the background losses are compensated above a pump power threshold of $ 130~\mathrm{mW/mm^{2}} $. This theoretical demonstration of a large gross gain per unit length for a \Nd doped active layer may justify further experimental work in order to achieve \Nd doped silicon based waveguide optical amplifier or laser.

In order to investigate the possibility of achieving larger gain further studies may be performed with other concentrations of rare earth and Si-ng, other rare earth and other pumping configurations. This method may be applied to study the electromagnetic fields and levels populations distribution in steady states of systems with other kind of emitters (quantum dots, quantum wells...) and in other configuration (VCELs, down-converting layers...).

\bigskip

\section*{Acknowledgments}
The authors are grateful to the French Nation Research Agency, which supported this work through the Nanoscience and Nanotechnology program (DAPHNES project ANR-08-NANO-005) and Centre de ressources informatiques de Haute-Normandie, (CRIHAN France) for computing facilities.

\end{document}